# Beam dynamics studies of the photo-injector in low-charge operation mode for the ERL test facility at IHEP

JIAO Yi (焦毅)，  XIAO Ou-Zheng (肖欧正)

*Institute of High Energy Physics, CAS, Beijing 100049, P.R. China*

**Abstract**: The energy recovery linac test facility (ERL-TF), a compact ERL-FEL (free electron laser) two-purpose machine, was proposed at the Institute of High Energy Physics, Beijing. As one important component of the TF, the photo-injector started with a photocathode direct-current gun has been designed. In this paper optimization of the injector beam dynamics in low-charge operation mode is performed with iterative scans using Impact-T. In addition, the dependencies between the optimized beam quality and the initial offset at cathode and element parameters are investigated. The tolerance of alignment and rotation errors is also analyzed.

## 1 Introduction

The energy recovery linac (ERL) and free electron laser (FEL) are considered to be candidates of the fourth generation light sources, and have received much attention worldwide. Since both of them are based on linac technologies, it is possible to combine FEL into an ERL facility, resulting in a compact two-purpose light source. A test facility, named energy recovery linac test facility (ERL-TF), was proposed at the Institute of High Energy Physics, Beijing, to verify this principle [1]. Physical design of the ERL-TF started a few years ago and is well in progress [2-4]. The layout and main parameters of the facility are presented in Fig. 1 and Table 1, respectively. Among the components of the facility, one extremely important device dominating the machine performance is the photo-injector. The injector, including a 500-kV photocathode direct-current (DC) gun equipped with a GaAs cathode, a 1.3 GHz normal conducting RF buncher, two solenoids, and two 2-cell superconducting RF cavities, was designed for the ERL-TF [2], with the layout shown in Fig. 2. With the initial parameters listed in Table 2, beam simulation of the designed injector was made for the high-charge operation mode (bunch charge 77 pC, rep. rate 130 MHz) with the ASTRA program [5], and finally an electron beam, with kinetic energy $E_k$ of 5 MeV, normalized emittance $\varepsilon_{n,x(y)}$ of 1.49 mm.mrad, rms bunch length $\sigma_z$ of 0.67 mm and rms energy spread $\sigma_\delta$ of 0.72%, was achieved at the end of the injector. In this paper, we optimize the beam dynamics of low-charge operation mode (bunch charge 7.7 pC, rep. rate 1.3 GHz) with iterative scans using the Impact-T program [6]. Thanks to the relatively weak space charge force, an electron beam with $\varepsilon_{n,x(y)}$ of 0.4 mm.mrad, $\sigma_z$ of 0.74 mm and $\sigma_\delta$ of 0.33% is obtained in the case of 0.5-mm incident laser rms transverse size. Moreover, the dependency of the beam quality on various variables, such as initial offset at cathode and element parameters, and the sensitivity of beam dynamics to element alignment and rotation errors are also investigated in this paper.

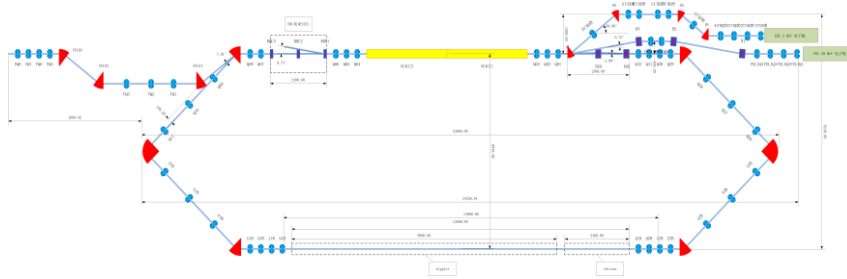

Fig. 1. Layout of the ERL test facility.

In the following, the detailed description of the optimization will be presented in Sec. 2, the dependency between the optimized result and the variables will be discussed in Sec. 3, and the error tolerance will be studied in Sec. 4.

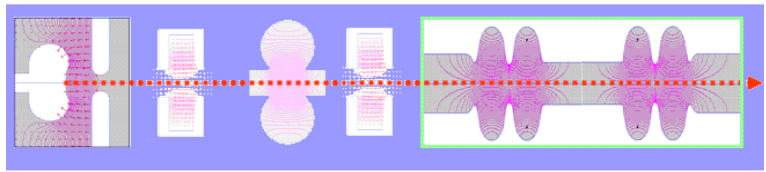

Fig. 2. Layout of the ERL-TF injector, consisting of, from left to right, DC-gun, the first solenoid, RF buncher, the second solenoid, and two 2-cell RF cavities.

Table 1. Main parameters of the ERL-TF at IHEP

| Parameter | Value |
| --- | --- |
| Beam energy (MeV) | 35 |
| Beam current (mA) | 10 |
| Bunch charge (pC) | 77 (or 7.7) |
| Normalized emittance (mm.mrad) | 1.0-2.0 |
| Rms bunch length (ps) | 2.0-4.0 |
| Rms energy spread (%) | 0.2-1.0 |
| Bunch frequency (MHz) | 130 (or 1300) |
| RF frequency (MHz) | 1300 |

Table 2. Initial parameters of the ERL-TF injector in Ref. [2]

| Parameter | Value |
| --- | --- |
| DC-gun voltage (kV) | 300-500 |
| Cathode material | GaAs |
| Driven laser | 2.3W, 532 nm |
| Laser rep. rate (MHz) | 130 (or 1300) |
| Laser trans. distr. | Round cross-section, uniform |
| Laser rms trans. size (mm) | 1.2 mm |
| Laser long. distr. | Beer-can with flat top of 20 ps, rise and fall time of 2 ps |
| E- ave. $E_k$ (eV) | 0.2 |

## 2 Beam dynamics optimization of the injector

The beam dynamics of the ERL-TF injector in low-charge operation mode is simulated and optimized with the Impact-T, a fully 3D program to track relativistic particles taking into account space charge force and short-range longitudinal and transverse wake-fields. The benchmark study between Impact-T and PARMELA showed good agreement in the simulation results [7], and this code has been used in the LCLS beam dynamics study [8] and an ERL injector optimization [9]. A parameter iterative scan program is developed with Matlab which starts several runs of tracking simultaneously. This code can finish the multi-variable scans, which usually contains a few hundred of runs, within an acceptable period of time (e.g. in 2 hours) on a desktop computer.

In the first stage of the study, we use the initial parameters and injector component fields the same (or as close as possible) as those in Ref. [2] except the bunch charge and repetition rate. The initial beam distribution for simulation is generated according to the laser parameters listed in Table 2, with round cross-section and longitudinal beer-can profile, as shown in Fig. 3. The initial beam has the same profile as the laser in $z$ dimension, while has a uniform kinetic energy distribution between 0 and 0.4 eV, with an average of 0.2 eV. The normalized emittance $\varepsilon_{n,x(y)}$ is given by

$$\varepsilon_{n,x(y)} = \sigma_{x(y)}\sqrt{\frac{k_B T_\perp}{m_e c^2}}, \tag{1}$$

where $\sigma_{x(y)} = 1.2$ mm, is the horizontal (vertical) rms beam size on the cathode, $m_e c^2$ is the electron rest energy, and $k_B T_\perp$ is the transverse beam thermal energy, which is found depending mainly on the incident laser wavelength [10],

$$k_B T_\perp (meV) = 309.2 - 0.3617\lambda(nm). \tag{2}$$

For the incident 532 nm laser, $k_B T_\perp = 116.8$ meV and $\varepsilon_{n,x(y)} = 0.57$ mm.mrad.

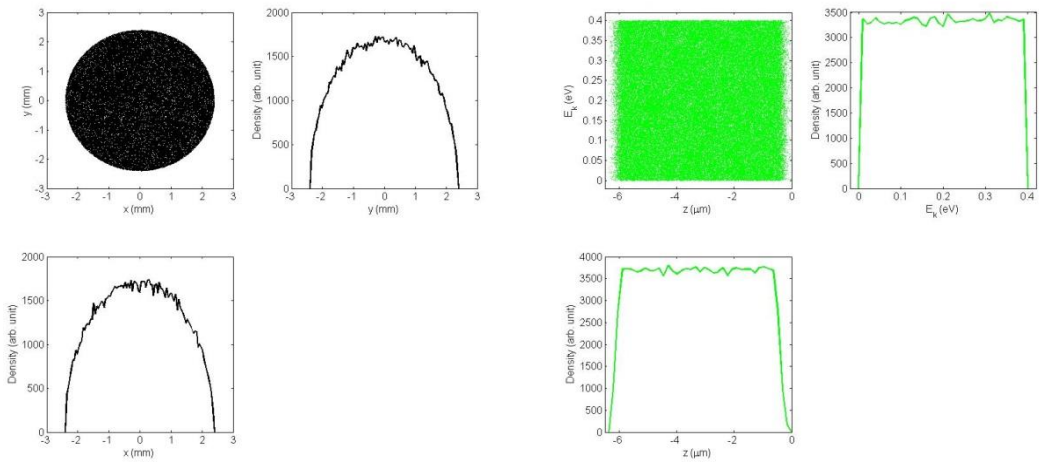

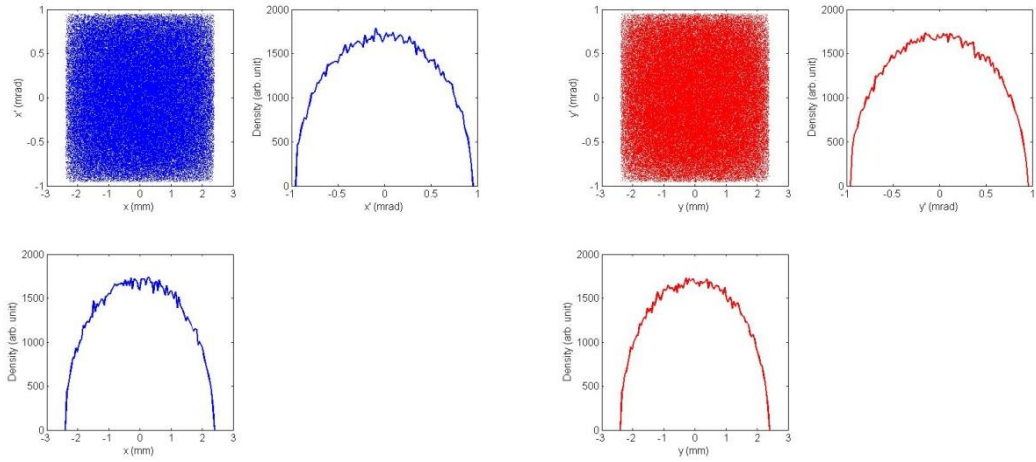

Fig. 3. Generated initial beam distribution in the phase space of $(x, y)$, $(z, E_k)$, $(x, x')$ and $(y, y')$ for Impact-T simulation.

With the generated initial beam distribution, twelve variables are iteratively scanned to search the optimal parameter setting that results in the lowest $\varepsilon_{n,x(y)}$, small $\sigma_\delta$, $\sigma_z$ of 2 ~ 4 ps, and $E_k$ of 5 MeV. The optimization starts with the scan of the buncher parameters to realize a $\sigma_z$ of 2 ~ 4 ps, then incudes the solenoid parameters in the scan to minimize the $\varepsilon_{n,x(y)}$ and the RF cavity parameters to optimize the $E_k$ as well as the $\sigma_\delta$ and $\sigma_z$, and finally ends with a global scan of all variables. Generally speaking, the variation of the solenoid position and strength contributes mainly to the emittance reduction, while slightly affects the $\sigma_z$ modulation. This is probably due to the medium roles of the space charge effect. More stringent squeeze of the transverse beam volume leads to stronger space charge force that will induce a change in $\sigma_z$. For the same reason the phase of the buncher field is the essential parameter modulating $\sigma_z$ (see Fig. 4), while also affecting the emittance reduction. The accelerating phases of the two 2-cell RF cavities are best to be separated from each other to achieve a high beam quality, such as moderate $\sigma_z$ and small $\sigma_\delta$.

The iteratively optimized results and the variables are shown in Table 3. With the 'Iterative 3' parameters, an electron beam with $E_k$ of 5 MeV, $\varepsilon_{n,x(y)}$ of 0.65 mm.mrad, $\sigma_z$ of 0.74 mm and $\sigma_\delta$ of 0.29% is achieved at the end of the injector. The field map of the elements along the beam line and the evolution of the beam parameters, such as $E_k$, $\varepsilon_{n,x(y)}$, $\sigma_{x(y)}$, $\sigma_z$ and $\sigma_\delta$, is presented in Fig. 5, and the final beam distribution is in Fig. 6.

It is realized that the laser transverse rms size is best to be smaller to make the simulation more close to the realistic condition. Thus, optimization for the case with a laser rms transverse size of 0.5 mm is made, with the result tabulated in Table 3 as well (see 'Result 2'). Even a smaller normalized emittance, $\varepsilon_{n,x(y)}$ = 0.40 mm.mrad, is obtained at the end of the injector. In the following, the dependency relationship and the error tolerance will be analyzed based on this case.

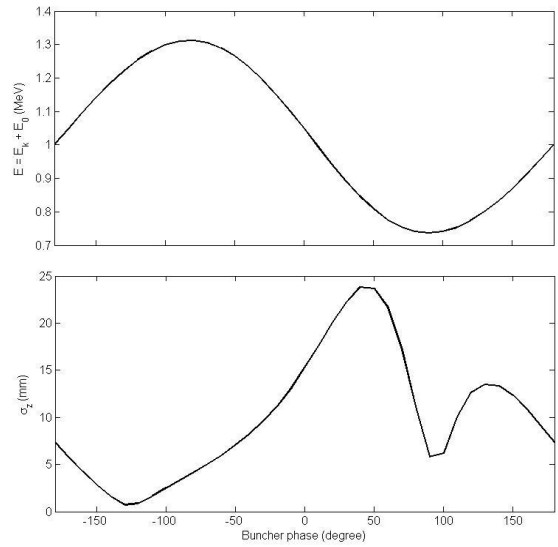

Fig. 4. Variations of the beam energy and the rms bunch length at the exit of the buncher with the phase of the buncher field.

Table 3. Iterative optimization results for the ERL-TF injector at IHEP

| Parameter | Iterative 1 | Iterative 2 | Iterative 3 | Result 2 |
|---|---|---|---|---|
| Laser rms tran. size (mm) | 1.2 | | | 0.5 |
| Final tran. Emittance (mm.mrad) | 0.71 | 0.65 | 0.65 | 0.40 |
| Final rms tran. size (mm) | 0.56 | 0.36 | 0.37 | 0.35 |
| Final rms bunch length (mm) | 0.62 | 0.71 | 0.74 | 0.74 |
| Final beam kinetic energy (MeV) | 4.98 | 4.91 | 5.00 | 5.00 |
| Final rms energy spread (%) | 0.55 | 0.74 | 0.29 | 0.33 |
| 1st solenoid position (m) | 0.35 | 0.42 | 0.41 | 0.43 |
| 1st solenoid peak field (Gauss) | 400 | 480 | 472 | 436 |
| Buncher position (m) | 0.8 | 0.8 | 0.8 | 0.8 |
| Buncher peak field (MV/m) | 4.62 | 4.62 | 4.62 | 4.62 |
| Buncher phase (degree) | -120 | -115 | -116 | -114 |
| 2nd solenoid position (m) | 1.1 | 1.15 | 1.14 | 1.11 |
| 2nd solenoid peak field (Gauss) | 480 | 480 | 488 | 548 |
| 1st cavity position (m) | 1.8 | 1.8 | 1.8 | 1.8 |
| 1st cavity peak field (MV/m) | 20 | 20 | 20 | 20 |
| 1st cavity phase (degree) | 20 | 20 | 22 | 23 |
| 2nd cavity position (m) | 2.65 | 2.65 | 2.65 | 2.65 |
| 2nd cavity peak field (MV/m) | 20 | 20 | 20 | 20 |
| 2nd cavity phase (degree) | 120 | 120 | 147 | 147.5 |
| Total length (m) | 3.3 | 3.3 | 3.3 | 3.3 |

## 3 Dependency relationship study

Recently significant progress was made in Cornell University on high-current operation from a photo-injector with a DC-gun [11]. One important technological improvement is to choose the active area off the cathode center, which helps avoiding the damage due to ion back-bombardment and hence providing good operational lifetime. To investigate the impact of the initial offset on the final beam quality, numbers of simulations with different initial offsets are performed. Since the beam distribution is no longer azimuthal symmetry with a nonzero offset, 3D space charge effects are turned on right at the beginning of the tracking. The result is shown in Fig. 7. It shows that a 5-mm offset from the cathode center does not lead to neither large difference between horizontal and vertical emittance nor large beam quality degradation. The emittance increases by about or more than 50%, but is still below 1 mm.mrad. It is interesting that the region with $\varepsilon_{n,x(y)} < 0.8$ mm.mard appears a diamond shape, instead of round. The underlying physics is not clear so far, and needs to be explored in the future.

Moreover, it is important to explore the dependency between the optimized beam quality and the element parameters, which will provide insight to the beam dynamics and help understand the influence of parameter fluctuation on the injector performance. To this end, each variable is varied around its optimal value ('Result 2' in Table 3), and the final beam parameters, such as $E_k$, $\varepsilon_{n,x(y)}$, $\sigma_z$, and $\sigma_\delta$, is recorded after simulation with Impact-T. The results are presented in Fig. 8, which shows several pieces of important information. First, it is verified that the normalized emittance is very close to, if not exactly on a (local) minimum. This is also confirmed by optimization with the multi-objective genetic algorithms which, however, will be addressed elsewhere. Secondly, the buncher phase that results in a minimum $\varepsilon_{n,x(y)}$ is not the same as that resulting in a minimum $\sigma_z$ or $\sigma_\delta$. The optimal buncher phase may vary with the beam quality requirement for different application purposes of the facility. Finally, the beam quality is very sensitive to the buncher and the RF cavity positions. However, it is realized that the change of a cavity position is equivalent to a change of the RF phase due to the fact that the particles will arrive the cavity earlier or later. Thus, the RF phase can be tuned accordingly to retrieve the optimal beam quality. For the 1.3 GHz RF buncher and the RF cavities, a 1-cm position deviation requires a change of 16 degree in RF phase.

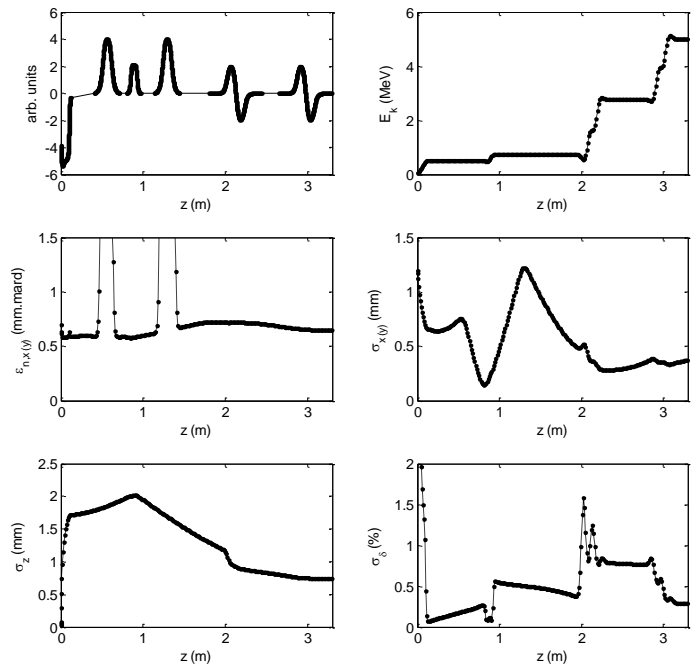

Fig. 5. The filed map of the elements and evolution of the beam parameters, such as such as $E_k$, $\varepsilon_{n,x(y)}$, $\sigma_{x(y)}$, $\sigma_z$ and $\sigma_\delta$, along the injector with the 'Iterative 3' parameters in Table 3.

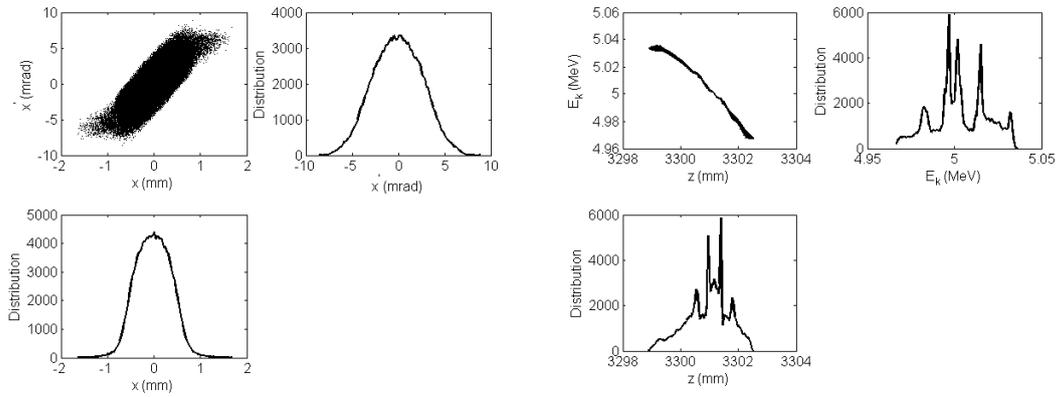

Fig.6. Beam distribution in the phase space of $(x, x')$ and $(z, E_k)$ at the end of the injector with the 'Iterative 3' parameters in Table 3.

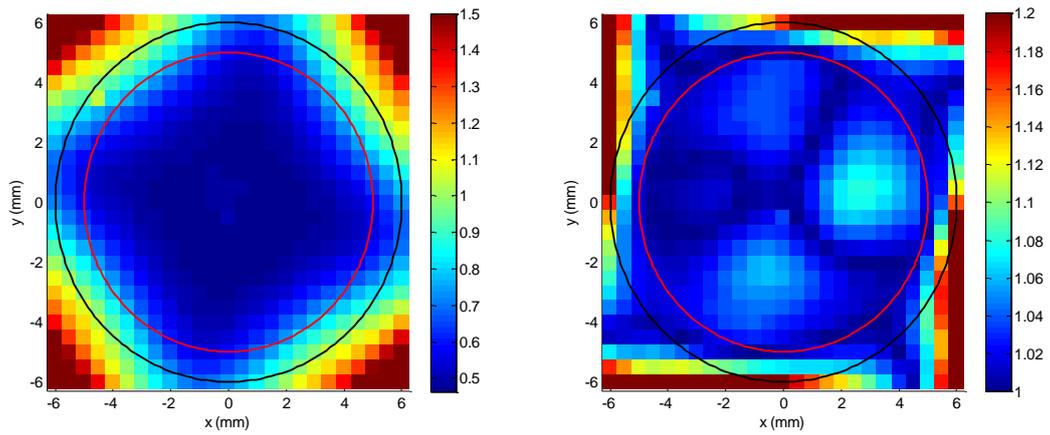

Fig. 7. (color online) Simulation data of $(\varepsilon_{n,x}^2/2 + \varepsilon_{n,y}^2/2)^{1/2}$ (left plot) and $max(\varepsilon_{n,x}/\varepsilon_{n,y}, \varepsilon_{n,y}/\varepsilon_{n,x})$ (right plot) with different initial offset at cathode. The inner (red) and the outer (black) circles represent initial offset of 5 mm and 6 mm from the center, respectively.

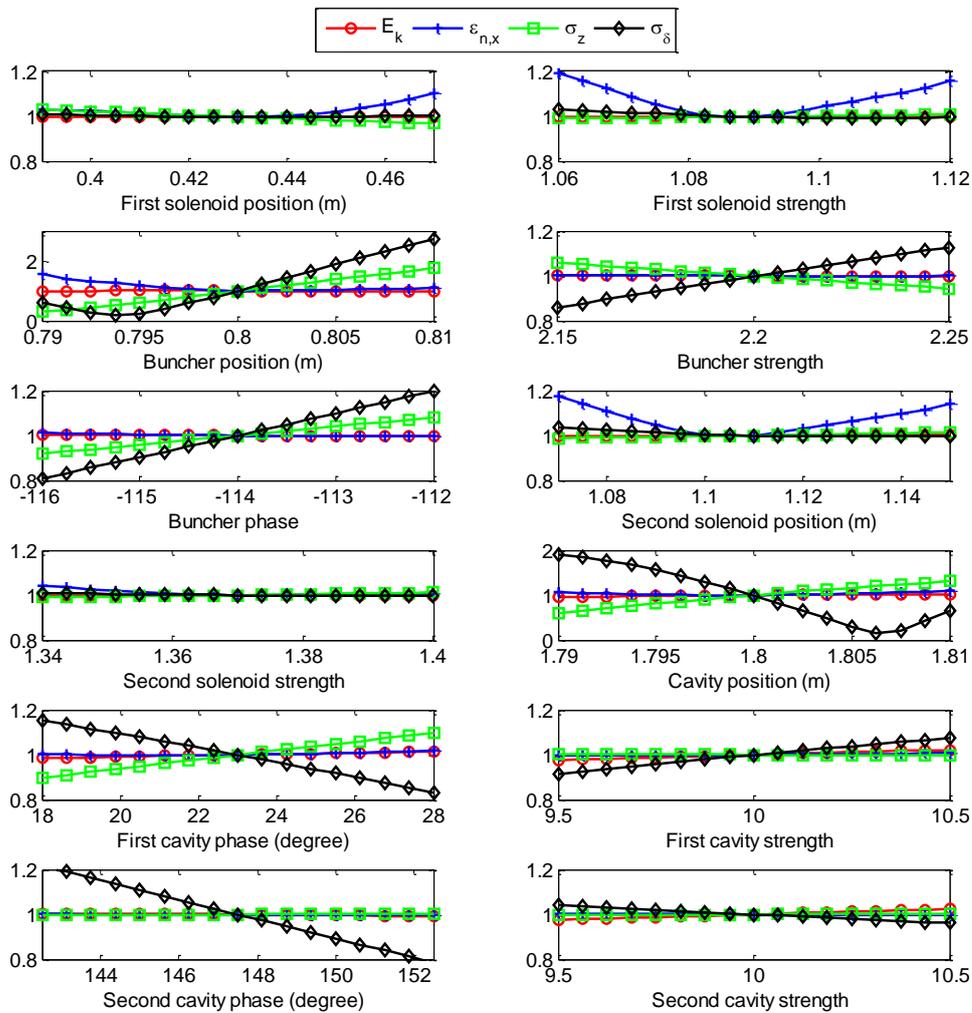

Fig. 8. Variations of the final beam parameters, such as $E_k$, $\varepsilon_{n,x}$, $\sigma_z$ and $\sigma_\delta$, with element parameters around the values of 'Result 2' in Table 3.

## 4 Error tolerance study

To ensure the feasibility of the optimized beam quality in a realistic condition, error tolerance study is necessary and tolerable magnitude of the errors should be determined. For the ERL-TF, both the alignment and rotation errors for each element are considered in the analysis. Presuming the alignment error and the rotation error have the same amplitude, we investigate the variation of the beam quality with error amplitude. For each specific error amplitude, 1000 random settings of the errors are added to each component, then tracking with 3D space charge forces is performed, and finally the beam parameters at the end of the injector are recorded. It is found that only the normalized emittance has evident increase due to errors. Therefore statistical analysis is performed only on emittance data. The statistics of the emittance growth in the case with 0.4 mm alignment and 0.4 mrad rotation errors is shown in Fig. 9. One can see that the growth rates $\Delta\varepsilon/\varepsilon$ spread out over a large range, with an average of 16.7% and a maximum of 60.2%. The variation of the average and maximum emittance growth rates with error amplitude is shown in Fig. 10.

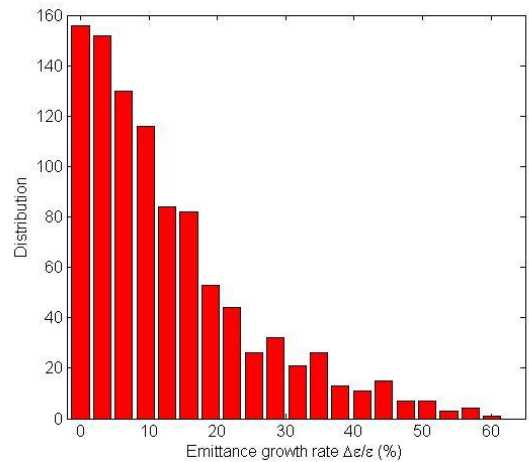

Fig. 9. Statistics of the emittance grow with 0.4 mm alignment and 0.4 mrad rotation errors.

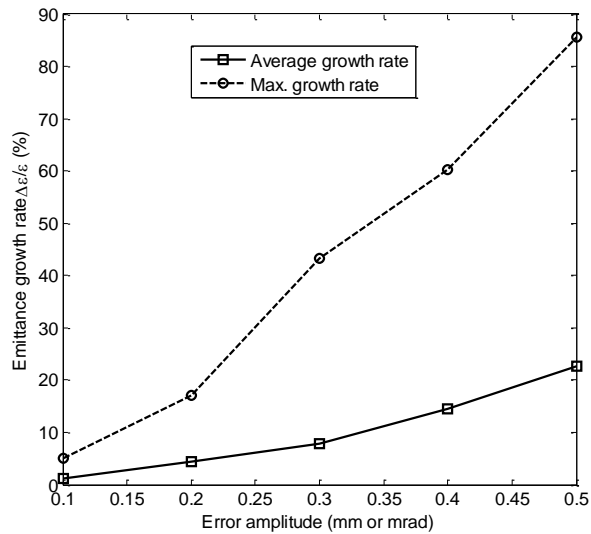

Fig. 10. Variation of the emittance growth rate with alignment and rotation error amplitude.

Furthermore, the contributions of different element and different error to emittance growth are analyzed. From the results shown in Table 4, only the alignment error of the two solenoids (especially the first solenoid) is the main source of the emittance growth. During construction of the injector, the alignment error of the solenoids should be strictly controlled to maintain a good machine performance. For a conservative estimation, to remain the emttance growth rate below 10%, the element alignment error of the solenoids must be smaller than 0.15 mm, while the other errors should be smaller than 0.3 mm or 0.3 mrad.

Table 4. Emittance growth due to different element and different error

| Element | Error | Ave. $\Delta\varepsilon/\varepsilon$ | Max. $\Delta\varepsilon/\varepsilon$ |
|---------|-------|---------|---------|
| 1st solenoid | Alignment 0.5 mm & rotation 0.5 mrad | 18.4% | 66.7% |
| | Alignment 0.5 mm | 16.4% | 61.1% |
| | rotation 0.5 mrad | 1.5% | 10.7% |
| Buncher | Alignment 0.5 mm & rotation 0.5 mrad | 0.41% | 1.28% |
| 2st solenoid | Alignment 0.5 mm & rotation 0.5 mrad | 7.9% | 38.1% |
| | Alignment 0.5 mm | 6.5% | 24.8% |
| | rotation 0.5 mrad | 0.28% | 1.39% |
| 1st RF cavity | Alignment 0.5 mm & rotation 0.5 mrad | 0.37% | 4.19% |
| 2nd RF cavity | Alignment 0.5 mm & rotation 0.5 mrad | 0.08% | 0.12% |

## 5 Conclusions

In this paper, we show the beam dynamics optimization of the ERL-TF injector in low-charge operation mode at IHEP with iterative scans using Impact-T program as well as the dependency analysis and the error tolerance study. It appears feasible to achieve a good beam quality at the end of the injector. The presented study is hoped to benefit future construction and commissioning of the ERL-TF facility.

## Acknowledgements

We thank the colleagues of the ERL-TF work group at IHEP for helpful discussions. Thanks also go to LIU Sheng-Guang for sharing the field data of the injector elements.